\documentclass[11pt]{article}
\usepackage{amsmath,graphicx}

\topmargin - 2cm
\oddsidemargin -
0.1cm \evensidemargin - 5cm

\begin{document}

\title{{\Large {\textbf{Diffusion of individual birds in starling flocks}}}}
\author{A. Cavagna$^{\sharp,\flat}$, S.~M. Duarte~Queir\'{o}s$^{\sharp}$, I. Giardina$^{\sharp,\flat}$, F. Stefanini$^{\dagger}$ and M. Viale$^{\sharp,\flat}$}
\date{}
\maketitle

\begin{center}
$^\sharp$ \emph{Istituto dei Sistemi Complessi, UOS Sapienza, CNR, via dei Taurini 19, 00185 Roma, Italy}\\[0pt]
$^{\flat}$ \emph{Dipartimento di Fisica, Universit\`a\ Sapienza, P.le Aldo Moro 2, 00185 Roma, Italy}\\[0pt]
$^{\dagger}$ \emph{Institute of Neuroinformatics, University of Zurich and ETH Zurich, Winterthurerstrasse, 190 CH-8057, Zurich, Switzerland}\\[0pt]
\end{center}

\begin{abstract}
Flocking is a paradigmatic example of collective animal behaviour, where global order emerges out of self-organization. Each individual has a tendency to align its flight direction with those of neighbours, and such a simple form of interaction produces a state of collective motion of the group.
As compared to other cases of collective ordering, a crucial feature of animal groups is that the interaction network is not fixed in time, as each individual moves and continuously changes its neighbours.
The possibility to exchange neighbours strongly enhances the stability of global ordering and the way information is propagated through the group. Here, we assess the relevance of this mechanism in large flocks of starlings ({\it Sturnus vulgaris}).
We find that birds move faster than Brownian walkers both with respect to the centre of mass of the flock, and with respect to each other. Moreover, this behaviour is strongly anisotropic with respect to the direction of motion of the flock. We also measure the amount of neighbours reshuffling and find that neighbours change in time exclusively as a consequence of the random fluctuations in the individual motion, so that no specific mechanism to keep one's neighbours seems to be enforced. On the contrary, our findings suggest that a more complex dynamical process occurs at the border of the flock.
\end{abstract}

\section{Introduction}

Self-organization and the spontaneous emergence of order in biological systems  does not come much more spectacular than in large flocks of starlings ({\it Sturnus vulgaris}). At dusk, huge flocks move above the roost, exhibiting beautiful collective patterns. There is no leader in the group and the collective movement is a unique consequence of local interactions between individuals \cite{camazine+al_01,couzin+krause_03}.

A central question in collective animal behaviour is to understand what are the interaction rules through which global coordination emerges. For a long time, due to the technical difficulties in reconstructing individual motion in large groups \cite{parrish+hammer_97}, data
have been scarce. More recently, though, a new generation of experimental studies, both in two and in three dimensions, have been
performed, establishing the basis for an empirically validated understanding of the interaction rules in collective animal behaviour
\cite{Theraulaz_2005, Couzin_2006,ballerini+al_08b,ballerini+al_08a,cavagna+al_08a,Lukeman_2010, Tunstrom_2011, Herbert-Read_2011}.
What these data show is that several traits of collective motion are well reproduced by relatively simple models based on local interaction rules \cite{aoki_82,vicsek+al_95,couzin+al_02,gregoire+chate_04,ginelli+chate_10,charlotte, bialek+al_12}. The fundamental ingredient shared by all models is the tendency of each individual to align to its neighbours. There is now a common consensus that this type of interaction is indeed a key aspect of collective motion in biology.

Alignment is a very important form of interaction in physics too: in ferromagnets the tendency of each spin to align to its neighbours gives rise to a spontaneous global magnetization, much as a flock of birds develops a spontaneous global velocity. However, in adopting such a minimalistic approach to the description of flocks, not only one makes a gross oversimplification of the individual entities (birds are not spins, of course), but also neglects a very fundamental difference between animal groups and spin systems: animals, unlike spins, move one with respect to another, so that the interaction network (i.e. who interacts with whom) changes in time. This crucial property of biological collective behaviour has a potentially large impact on how information propagates throughout the group.

There are indeed two mechanisms that contribute to the emergence of global coordination. The first one is the direct alignment of one individual with its interacting neighbours; from neighbour to neighbour local ordering spreads over the interaction network to the whole group. This mechanism works even if individuals do not move one with respect to another, like spins sitting on the sites of a crystalline lattice. The second mechanism, on the contrary, is intrinsically related to motion: when individuals move, two animals that were not directly interacting at a given time, may become proximate neighbours and interact at a later time, so that information is more efficiently propagated throughout the group. It has been hypothesized that this mechanism reinforces correlations between individuals, strongly enhancing global ordering \cite{toner+tu_95,toner+tu_98,jadbabaie}.

This extra ingredient of collective animal behaviour implies that we cannot simply investigate {\it static} aspects of the interaction network (like, for example, the number of interacting neighbours \cite{ballerini+al_08a}), but we need to get information about the {\it dynamical} evolution of the interaction network. A first step in this direction is to study how individual animals move and rearrange within the group. This is what we do here for flocks of starlings in the field.

There are two other important reasons why it is relevant to have information about the relative dynamics of individuals.
It has been found in \cite{ballerini+al_08a}, and later confirmed in \cite{bialek+al_12}, that starlings in a flock interact with a fixed number of neighbours, rather than with all neighbours within a fixed metric radius. This number is approximately seven. A natural question is: what is the permanence in time of these
seven individuals? Do they change uniquely due to the relative motion between individuals? Or is there any kind of relationship between interacting neighbours that keeps them together longer?

A second question regards the border of the flock. Birds at the border are more exposed to predation than those at the interior. Former studies showed that the density of the flock at the border is larger than at the interior, probably as a consequence of the fact that border birds `push' towards the inner part of the flock to get in \cite{ballerini+al_08a}. Is there a border turnover? If yes, how fast is it?

To quantify how individuals move through the group we use a statistical perspective and adopt the powerful approach of diffusion processes \cite{gardiner,berg}. To study diffusion one needs not only the positions and velocities of the birds, but the full individual trajectories. Individual
tracking is a further level of difficulty with respect to static 3D reconstruction (see Methods) and a good performance is strictly related to having
fast enough cameras and a large memory, in order to record long events. Even though this was not quite the case in our past experiments
\cite{ballerini+al_08a, ballerini+al_08b, cavagna+al_10}, we succeed for a few flocking events, and for not-too-long a time interval, in retrieving
a reasonable percentage of trajectories, with a sampling rate of $10$ frames per second (see Table 1). Using these trajectories, we compute the diffusion properties of individuals with respect to the center of mass and to neighbours. Moreover, we study the neighbours reshuffling rate and show how it is connected to the diffusion properties of individuals. Finally, we study  the dynamics at the border of the flock.

\section{Results}
\subsection{Quantifying individual motion through diffusion}
Why do individuals move through the flock and exchange positions? If each bird {\it exactly} chose the direction of motion and speed of its neighbours, one would get a perfect flock where every individual keeps following the same direction as others. Relative positions would remain the same, defining an interaction network (who interacts with whom)  that is fixed in time.  But  imitation and mutual alignment are never complete,  there is always an amount of uncertainty or arbitrariness in the individual choices. As a consequence, flight directions between neighbours are very similar, but not identical, differing by small `random' fluctuations that flocking models usually describe through a stochastic noise term. Time after time these fluctuations accumulate, determining a departure of the individual trajectories and a reshuffling of neighbourhood relationships. To describe such a process, it is useful to consider first the case where social forces are absent and individuals merely follow random moves. This is the renowned case of Brownian motion (where - originally - the random walkers were particles instead of birds). To quantify how much the Brownian walkers move in time, one can look at the average mean-square displacement as a function of time, i.e. at the average amount of distance travelled in a time $t$:
\begin{equation}
\delta R^2 \left( t\right) \equiv \frac{1}{T-t}\frac{1}{N}%
\sum_{t_{0}=0}^{T-t-1}\sum\limits_{i=1}^{N}\left[ \vec{R}_{i}\left(
t_{0}+t\right) -\vec{R}_{i}\left( t_{0}\right) \right]^2 ,
\label{BM}
\end{equation}
where ${\vec R}_i(t)$ indicates the position of bird/particle $i$ at time $t$, and where we have averaged over all  $N$ individuals in the group and over all time lags of duration $t$ in the interval $[0,T]$. For Brownian motion the mean-square displacement grows linearly with time \cite{gardiner}, i.e. $\delta R^2 \left( t\right) \propto t$, indicating that in their random wandering walkers depart increasingly from their origin. This behaviour, which is referred to as standard (or normal) diffusion, is rather robust and usually persists even in presence of external forces or interactions between individuals. In some cases, however, such forces can enhance/deplete in a non-trivial way the effect of noise, leading to different diffusion laws.
 The majority of natural processes is well-described by a power-law dependence,
\begin{equation}
\delta R^2
\left( t\right) =D \, t^{\alpha },
\label{power-law}
\end{equation}
where $\alpha$  -  the diffusion exponent - falls between $0$ and $2$, and $D$ is the diffusion coefficient. The case $\alpha =1 $ corresponds to Brownian motion and  to normal diffusion.  When $\alpha >1$ particles move/diffuse faster, and this is why this case is indicated as super-diffusive (the special case $\alpha =2$ corresponding to ballistic diffusion).
Finally, we note that  although for very long times the type of diffusion is characterized by the value of the exponent $\alpha$, for finite times even the value of the coefficient $D$ plays a key role, larger values of $D$ corresponding to more mobile particles/individuals.

\subsection{Diffusion in the centre of mass reference frame}

Coming back to flocks, our aim is now to use the above definitions to quantify how much individuals move through the group and one with respect to the other. Since flocks are strongly ordered, each bird moves predominantly in the same direction as the whole group. This contribution to individual motion is common to all birds and, if deviations were absent, would entail a fixed network of reciprocal positions. We are rather interested in what makes this network changing in time. Therefore, we need to take away this global component and focus on individual movements with respect to the flock's motion. This can be done by considering the birds movements in the centre of mass reference frame: at each instant of time the coordinates of an individual in this reference frame define its location inside the flock and, correspondingly, diffusion describes how much a bird has changed its position within the group (while at the same time co-moving with it). To visualize this point, in Fig.~1 we show a couple of trajectories of neighbouring birds, both in the cameras reference frame and in the flock's centre of mass reference frame.
We notice that the centre of mass reference frame  closely resembles the subjective perception individuals have of collective motion when flying together.
Birds individual velocities are in fact very close to the centre of mass one (being the flock very polarized), therefore the centre of mass frame is very similar to a frame co-moving with the birds themselves. An even more faithful representation of the individual perception (for a given bird) is provided by the mutual diffusion setting (see next section).

\begin{figure}[h]
\begin{center}
\includegraphics[width=0.75\columnwidth,angle=0]{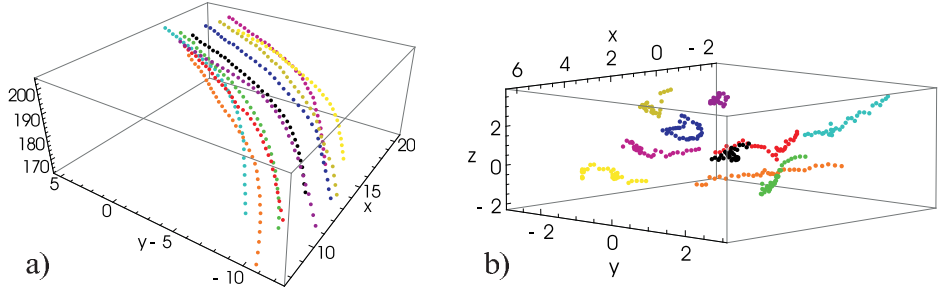}
\end{center}
\caption{Left: 3D reconstruction of some trajectories of flock 69--10 (1124 individuals) in the laboratory reference frame. Right:
The same trajectories in the centre of mass reference frame. All the axes are in meters.}
\end{figure}

To quantify diffusion behaviour in the centre of mass reference frame we consider the mean-square displacement as in Eq.~(\ref{BM}), but where coordinates are expressed in the centre of mass frame, e.g.
\begin{equation}
\delta r^2 \left( t\right) \equiv \frac{1}{T-t}\frac{1}{N}%
\sum_{t_{0}=0}^{T-t-1}\sum\limits_{i=1}^{N}\left[ \vec{r}_{i}\left(
t_{0}+t\right) -\vec{r}_{i}\left( t_{0}\right) \right]^2 ,
\label{msd}
\end{equation}
where $\vec{R}_{CM}(t)$ indicates the position of the centre of mass of the flock at time $t$, and $\vec{r}_{i}\left( t\right) =\vec{R}_{i}\left( t\right) -\vec{R}_{CM}\left( t\right) $ therefore represents the position of bird $i$ in the center of mass reference frame.  Besides,  $N$ is the number of birds in the flock and  $T$  the length of the time series.

\begin{figure}[h]
\begin{center}
\includegraphics[width=0.75\columnwidth,angle=0]{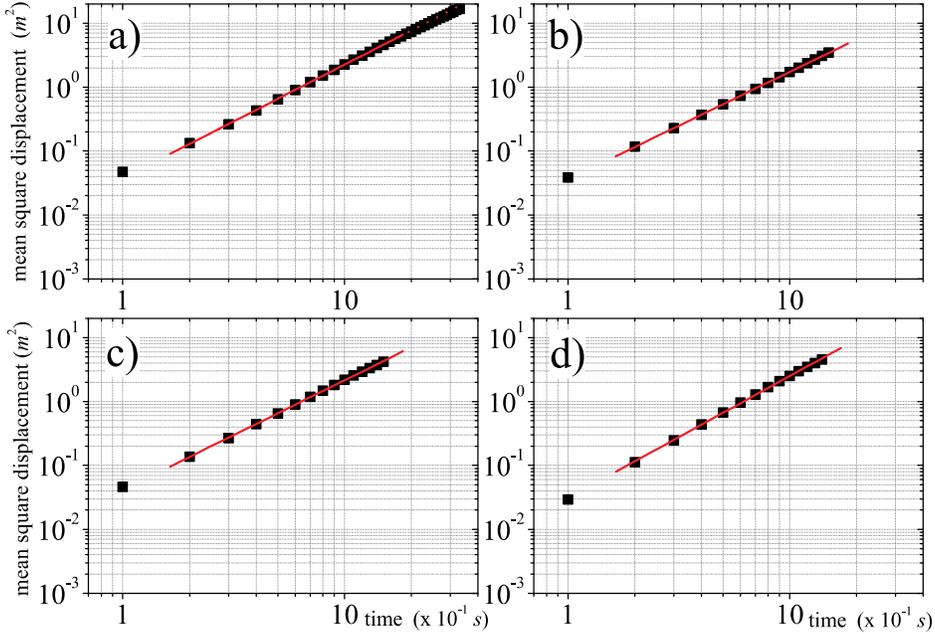}
\end{center}
\caption{Mean-square displacement in the centre of mass reference frame, for $4$ different flocking events:
a) 69-10; b) 48-17; c) 49-05; d) 28-10. Values of diffusion exponent and diffusion constant for each flock are: $\alpha=1.77\pm 0.02$, $D=3.8\pm0.02$ (a); $\alpha=1.73\pm 0.03$, $D=3.5\pm0.03$ (b); $\alpha=1.71\pm 0.02$,$D=3.9\pm0.03$ (c); $\alpha=1.83\pm 0.01$, $D=3.8\pm0.01$ (d).
}
\end{figure}

Using 3D trajectories of individual birds in starling flocks, we computed the mean-square displacement following Eq.~(\ref{msd})  for six flocking
events (see Methods).
We find that diffusion of birds satisfies quite well the time-dependence described by Eq.\ (\ref{power-law}), with an
exponent that is systematically larger than 1, \emph{i.e.}, birds perform {\it super-diffusive} motion in the center of mass reference frame. In  Fig.~2  we present the data of 4 flocks, but results are similar in the other analyzed flocks (see Table 1). Averaging the diffusion exponent over all the analyzed events we get,
\begin{equation}
\alpha =1.73\pm 0.07
\quad
\quad
,
\quad
\quad
D =0.036\pm0.004 \ .
\label{gof}
\end{equation}

\begin{table}[tbh]
\label{table}
\begin{center}
\begin{tabular}{|llllllll|}
\hline
Event & $\ N$ & $\ T\,\left( s\right) \ $ & $\ N_{LL}\ $ & $\ \alpha $ & $\
\alpha _{ij}$ & $\ D\,\left( \times \,10^{-2}\right) \ $ & $\ D_{m}\,\left(
\times \,10^{-2}\right) \ $ \\ \hline
28-10 & \multicolumn{1}{c}{$\ 1246\ $} & \multicolumn{1}{c}{$\ 1.5$} &
\multicolumn{1}{c}{$\ 785\ $} & \multicolumn{1}{c}{$\ 1.83\pm 0.01\ $} &
\multicolumn{1}{c}{$\ 1.88\pm 0.02\ $} & \multicolumn{1}{c}{$3.8\pm 0.1$} &
\multicolumn{1}{c|}{$0.37\pm 0.04$} \\ \hline
48-17 & \multicolumn{1}{c}{$\ 871\ $} & \multicolumn{1}{c}{$\ 1.6$} &
\multicolumn{1}{c}{$\ 350\ $} & \multicolumn{1}{c}{$\ 1.73\pm 0.03\ $} &
\multicolumn{1}{c}{$1.48\pm 0.02$} & \multicolumn{1}{c}{$3.5\pm 0.3$} &
\multicolumn{1}{c|}{$1.7\pm 0.03$} \\ \hline
49-05 & \multicolumn{1}{c}{$\ 797\ $} & \multicolumn{1}{c}{$\ 1.6$} &
\multicolumn{1}{c}{$\ 146\ $} & \multicolumn{1}{c}{$\ 1.71\pm 0.02\ $} &
\multicolumn{1}{c}{$1.50\pm 0.02$} & \multicolumn{1}{c}{$3.9\pm 0.3$} &
\multicolumn{1}{c|}{$0.77\pm 0.06$} \\ \hline
58-06 & \multicolumn{1}{c}{$\ 442\ $} & \multicolumn{1}{c}{$\ 3.1$} &
\multicolumn{1}{c}{$\ 140\ $} & \multicolumn{1}{c}{$\ 1.69\pm 0.01\ $} &
\multicolumn{1}{c}{$1.55\pm 0.02$} & \multicolumn{1}{c}{$3.6\pm 0.2$} &
\multicolumn{1}{c|}{$1.1\pm 0.04$} \\ \hline
69-09 & \multicolumn{1}{c}{$\ 239\ $} & \multicolumn{1}{c}{$\ 4.6$} &
\multicolumn{1}{c}{$\ 62\ $} & \multicolumn{1}{c}{$\ 1.64\pm 0.02\ $} &
\multicolumn{1}{c}{$1.32\pm 0.01$} & \multicolumn{1}{c}{$4.1\pm 0.3$} &
\multicolumn{1}{c|}{$1.7\pm 0.04$} \\ \hline
69-10 & \multicolumn{1}{c}{$\ 1129\ $} & \multicolumn{1}{c}{$\ 3.4$} &
\multicolumn{1}{c}{$\ 500\ $} & \multicolumn{1}{c}{$\ 1.77\pm 0.02\ $} &
\multicolumn{1}{c}{$1.72\pm 0.02$} & \multicolumn{1}{c}{$3.8\pm 0.2$} &
\multicolumn{1}{c|}{$0.89\pm 0.05$} \\ \hline
\end{tabular}%
\caption{Table of the analyzed flocks.
The number of birds $N$ is the number of individuals for which we obtained a 3D reconstruction of positions in space (average over all frames). The duration $T$  of the event is measured in seconds = number of frames $\times 10^{-1}$s. $N_{LL}$ indicates the number of retrieved trajectories that are as long as the entire time inerval $T$. The last 4 columns give the values of diffusion and mutual diffusion parameters.}
\end{center}
\end{table}

\subsection{Mutual diffusion}
 The results discussed above indicate that individuals move within the group faster than Brownian walkers. This super-diffusive behavior is probably the consequence of the interacting nature of collective motion, which give rise to strong correlations between birds' flight directions. Velocity correlations were first studied for bird flocks in \cite{cavagna+al_10}, where it was found that there are large correlated domains of birds with highly aligned velocities fluctuations.  This means that if a bird is moving in a certain direction with respect to the center of mass, its neighbours will move along similar directions \cite{cavagna+al_10}. This fact suggests that diffusive displacement of a bird with respect with its neighbours should be smaller than with respect to the centre of mass. Is it so?

We can answer this question by calculating how much individuals in the flock move with respect to one another.
We define an expression very similar to the \eqref{msd}, but in which {\it mutual} mean square displacement of birds $i$ with respect to
its nearest neighbour $j$ at time $t_0$,  is considered,
\begin{equation}
\delta r_{m}^2\left( t\right) \equiv \frac{1}{T-t}\frac{1}{N}%
\sum_{t_{0}=0}^{T-t-1}\sum\limits_{i=1}^{N}
\left[ |\vec{s}_{ij}\left(t_{0}+t\right)| - |\vec{s}_{ij}\left( t_{0}\right)| \right] ^{2},
\label{mutdiff}
\end{equation}
where $\vec{s}_{ij}\left( t\right) \equiv $ $\vec{r}%
_{i}\left( t\right) -$ $\vec{r}_{j}\left( t\right) $ is the
position of bird $j$ (the nearest neighbour of $i$ at time $t_{0}$) in the reference frame of $i$.
Also for mutual diffusion we find a power law behavior,
\begin{equation}
{\delta }r_{m}^2\left( t\right) = D_{m} \, t ^{\alpha_{m}}  \ .
\label{rantolo}
\end{equation}
Averaging over all flocks, we obtain (for individual flocks' value consult Table~1),
\begin{equation}
\alpha_{m}=1.58\pm 0.2
\quad
\quad
,
\quad
\quad
D_{m} = 0.011 \pm 0.005  \ .
\label{lobello}
\end{equation}
The representation of the time dependence of $\delta r_{m}^2(t)$ for the same flocks of Fig.~2 can be found in Fig.~5 in the Supplementary Information (SI). From a comparison between the average parameters in \eqref{gof} and \eqref{lobello} as well as the figures we have just mentioned, we can see that even though both diffusion and mutual diffusion have an exponent larger than $1$, mutual diffusion is suppressed with respect to diffusion in the centre of mass. The available statistics does not allow to conclude that the exponents are different (see SI), however in the intermediate time regime we are dealing with, this suppression is clear, especially looking at the diffusion coefficients.
Mutual diffusion describes how, on average, an individual bird perceives the motion of its neighbours relative to its own. As we shall see, it is a crucial information to understand how neighbours reshuffling occurs in a flock.

\subsection{Anisotropic diffusion}

To push further our analysis, we can ask whether diffusion and relative motion occur isotropically or whether, on the contrary, privileged directions exist.
The simplest way to probe the existence of privileged directions is to consider a matrix generalization of  Eq.~(\ref{msd}) (see SI for details). If we diagonalize this matrix, the diagonal elements automatically provide the mean-square displacement along the principal axes of diffusion (see Fig.~7 for an example in $4$ flocks). We can then compute, for each flock, the diffusion exponent and the diffusion coefficient along each axis.
What we find is that diffusion is strongly anisotropic, occurring more strongly along certain directions (corresponding to larger diffusion exponents) than others.  More precisely, the average diffusion exponents and coefficients along the three principal axes are,
\begin{eqnarray}
\alpha_1 =1.78\pm 0.08
\quad
\quad
D _1 = 0.021\pm0.003
\nonumber
\\
\alpha_2 =1.63\pm 0.04
\quad
\quad
D _2 =0.008 \pm 0.004
\nonumber
\\
\alpha_3 =1.44\pm 0.12
\quad
\quad
D _3 = 0.004 \pm 0.004
\nonumber
\end{eqnarray}

Previous studies~\cite{ballerini+al_08b} showed that flocks tend to fly parallel to the ground, and therefore orthogonal to gravity.
It is therefore natural to analyze the relation between the three principal axes of diffusion and the directions in space
that are naturally relevant for a cruising flock, namely the direction of motion and gravity. To investigate this point, we computed the average (in time) scalar product of the three normalized eigenvectors of diffusion, ${\bf u}_1$,  ${\bf u}_2$, ${\bf u}_3$, with the normalized vectors of the flock velocity, ${\bf u}_V$ and of gravity, ${\bf u}_G$. The results are the following,
\begin{eqnarray}
{\bf u}_1 \cdot {\bf u}_V &=& 0.11 \pm 0.04
\quad
\quad
,
\quad
\quad
{\bf u}_1 \cdot {\bf u}_G = 0.27 \pm 0.07
\nonumber
\\
{\bf u}_2 \cdot {\bf u}_V &=&0.90   \pm 0.03
\quad
\quad
,
\quad
\quad
{\bf u}_2 \cdot {\bf u}_G = 0.48 \pm 0.19
\nonumber
\\
{\bf u}_3 \cdot {\bf u}_V &=& 0.41 \pm  0.12
\quad
\quad
,
\quad
\quad
{\bf u}_3 \cdot {\bf u}_G = 0.83 \pm 0.11
\nonumber
\label{asino}
\end{eqnarray}
From these data we see that gravity, ${\bf u}_G$, has very high alignment with the direction of lowest diffusion, ${\bf u}_3$, while it has very low alignment with the
direction of the largest diffusion, ${\bf u}_1$. The direction of global motion, ${\bf u}_V$, has very high alignment with the second smaller diffusion direction, ${\bf u}_2$, while  (as gravity) it has minimal alignment with the direction of largest diffusion, ${\bf u}_1$. We conclude that diffusion is suppressed along gravity and direction
of motion, while the axis of maximal diffusion, ${\bf u}_1$, is  approximately perpendicular  to both group velocity and gravity, and therefore it roughly coincides with the wings axis.

The fact that diffusion along gravity is very limited is perhaps unsurprising, because of the energy expenditure that vertical motion requires.
On the other hand, the higher weight of diffusion along the wings direction vs. the velocity direction is less obvious on a purely biological basis.
As we shall see in the Discussion, though, previous theoretical investigations indeed predicted that diffusion in flocking had to be much stronger along a direction orthogonal to the direction of motion, which is exactly what we observe here.

\subsection{Neighbours reshuffling}

A crucial consequence of motion and of mutual diffusion is that individuals may change their neighbours in time.
Let us consider a (focal) bird $i$ at an initial time $t_{0}$ and its $M$ nearest neighbours. After a time $t$, some of these $M$ birds will not
belong to the set of neighbours of $i$ anymore. To monitor how the neighbourhood changes, we can calculate the percentage of individuals
that remain within the set of the $M$ nearest neighbours of $i$ after a time $t$.  Let us therefore define the {\it neighbours overlap} as,
\begin{equation}
Q_{M}(t) = \frac{1}{N}\sum_i\frac{M_i\left( t\right) }{M},
\label{overlap}
\end{equation}
where $M_i\left( t\right) $ is the number of birds that are among the $M$ nearest
neighbours of bird $i$ at both $t_{0}$ and $t+t_{0}$. The average runs over all the birds in the flock and over all initial times $t_0$.

In Fig.~3, we show the evolution of the overlap, $Q_{M}\left( t\right) $, as a function of the time $t$ and
number of neighbours $M$. Clearly, if we set $M=N$, \emph{i.e.}, if we choose a neighbourhood  as large as the whole flock, the overlap
remains by definition constant and equal to 1. When $M<N$, we see that the overlap smoothly decreases in time due to birds motion.
We conclude that neighbours reshuffling {\it does} happen, even for very close neighbours. This implies that the interaction network is changing
in time and that there is no indication of a preferred structure of neighbours in the flock. We also notice, however, that the process or reshuffling
the neighbours occurs on a timescale of a few seconds, which is rather long. We will analyze the implications of this fact in the Discussion.

\begin{figure}[h!]
\begin{center}
\includegraphics[width=0.75\columnwidth,angle=0]{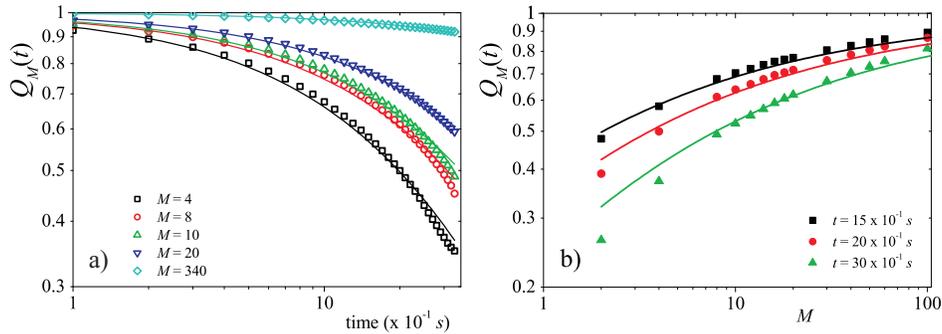}
\end{center}
\caption{Left: neighbours overlap $Q_{M}(t)$ vs $t$. Right: $Q_{M}(t)$ vs $M$.
Full lines represent Eq. (\protect\ref{overlappo}) with $c=0.048$ (fitted value), while $\hat d=2.3$
and $\protect\alpha =1.7$ have the values predicted by the geometrical argument described in the SI. Data are for flock 69-10.
}
\end{figure}

Interestingly, it is possible to explain the behavior of the cluster overlap purely in terms of the diffusion properties described in the previous sections.
The basic idea is simple: consider a focal bird, and its neighbourhood of $M$ birds. We ask how many neighbours the focal bird can lose in a time $t$. The most at risk are those in the outer edge of the neighbourhood. We make the very crude approximation that in a time $t$ the outer birds will have traveled a distance $l \sim \sqrt{D_{m} t^{\alpha_{m}}}$, which is a sort of deterministic interpretation of mutual diffusion equation \eqref{rantolo}. In this case, the number of lost neighbours will be of the order $\rho R^2 l$, where $R$ is the radius of the neighbourhood, which is connected to $M$ by the simple relation, $M\sim \rho R^3$. Using this argument we finally get (see SI for the details),
\begin{equation}
Q_{M}(t) =\left( 1+c\; \frac{t^{\alpha_{m}/2}}{M^{1/\hat d}}\right) ^{-\hat d}  \  ,
\label{overlappo}
\end{equation}
where $c$ is a constant related to the flock density $\rho$ and to the mutual diffusion coefficient $D_{ij}$ (see the SI).
For infinite and homogenous flocks $\hat d$ coincides with the space-dimension, $\hat d = 3$, whereas for finite flocks, due to
the presence of the border, we have an effective dimension $\hat d < 3$. The value of $\hat d$ can be fixed by making
a power law fit to the formula, $M\sim a R^{\hat d}$ (see Fig.~6 in the SI). We find $\hat d = 2.3$.

Using the value of $\alpha_{m}$ obtained in the previous sub-section, we get a very good agreement with the data (Fig.~3), both for what concerns the dependence of $Q$ on $t$ and on $M$.
Such agreement indicates that neighbours reshuffling is entirely ruled by diffusion: there seems to be no {\it ad hoc} mechanism used by birds to pick up their neighbours, nor any specific attempt to keep them fixed in time. Rather, neighbours reshuffling is simply the result of diffusion taking its course, so that at each instant of time each bird is interacting with whatever birds have been brought there by their superdiffusive wandering throughout the flock.

\subsection{Permanence on the border}

Because of the attacks of predators and of possible interactions with other external perturbations,
birds at the border of the flock might exhibit specific dynamical properties \cite{hamilton_71}.
To investigate this issue, we calculate the border survival probability, $P(t) $, defined as the probability that a bird initially at the border remains on the border for a time greater than $t$. (For a precise definition of the flock's border see SI). The data for $P(t)$ are shown in Fig.~4 for four different flocks.

\begin{figure}[h!]
\begin{center}
\includegraphics[width=0.75\columnwidth,angle=0]{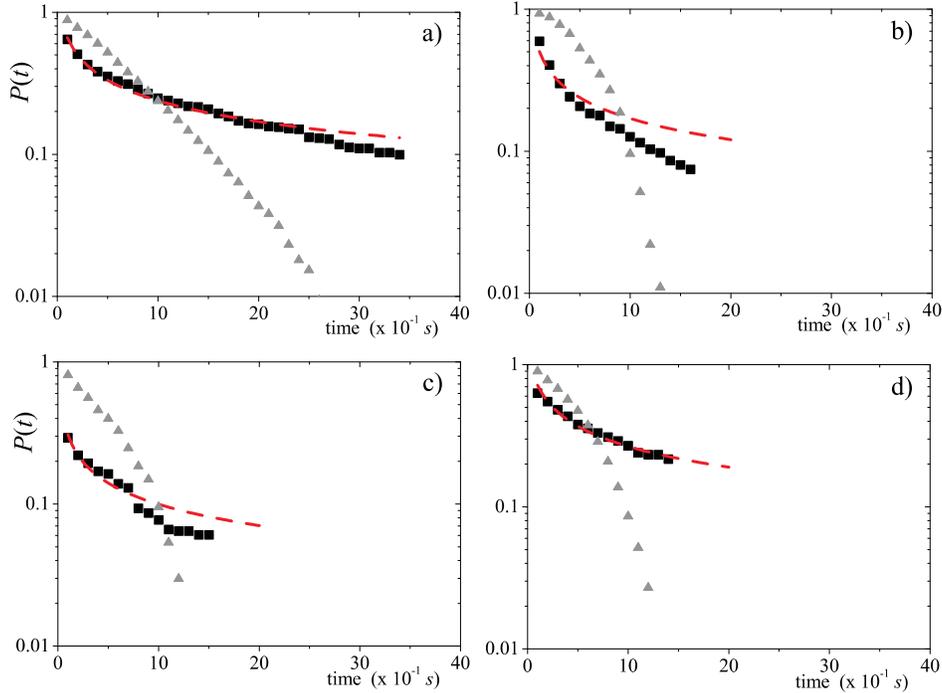}
\end{center}
\caption{Border survival probability  $P(t)$ (full black symbols) for 4 different flocking events : a) 69-10; b) 48-17; c)
49-05; d) 28-10. The dashed red line corresponds to a fit of the data using a Brownian functional form. For each flock we also report (light grey symbols) the survival probability for internal individuals.}
\end{figure}

Given our success in explaining neighbours reshuffling by purely  using the diffusion properties, it is interesting to ask whether the border survival probability too is ruled simply by diffusion or whether  there is some extra dynamical ingredient ruling the way birds remain on the border.
We may start saying that once a bird has travelled more than the average distance $l_\mathrm{B}$ between border and first internal nearest neighbour, it has left the border. If we use the same crude approximation as for neighbours reshuffling, namely that in a time $t$ a bird travels on average a distance $\sqrt{D t^\alpha}$, we can get an estimate of the time scale birds remain on the border,
\begin{equation}
\tau_\mathrm{diff}= (l_\mathrm{B}^2/D)^{1/\alpha}  \ .
\label{taudiff}
\end{equation}
Using for $l_\mathrm{B}, \alpha$ and $D$ the measured values for flock 69-10, this gives $\tau_\mathrm{diff}=0.8 s$. On the other hand, if we look at the data in Fig.~4 we can see that after a fast and short initial decay, the curve exhibits a rather long tail, indicating that persistence on the border can in fact be much longer.  A simple exponential fit of flock 69-10 gives indeed a time scale $\tau=2.5 s$, a factor $3$ larger than what mere diffusion predicts. A similar underestimation occurs for the other flocks. It seems that a naive diffusion argument does not work and that individuals at the border tend to exchange positions with neighbours less than what internal individuals do.

The discrepancy between internal and border dynamics is confirmed by comparing the border survival probability with the analogous survival probability for internal birds. In Fig.~4 together with the border $P(t)$ we also plot the probability that in a time $t$ an internal bird remains within a distance $l_\mathrm B$ from its initial position in the centre of mass reference frame (i.e. the probability that a positional swap with a neighbour does not occur). This internal survival probability decays much faster than the border one. Analytic computations allow to compute rigourous bounds for the survival probability of a super-diffusive walker with diffusion exponent $\alpha$ (see SI and \cite{monrad+rootzen_95}). The empirical $P(t)$ for internal birds is fully consistent with these predictions, confirming once again that mutual rearrangements of internal individuals can be explained in terms of a diffusion mechanism. On the contrary, the same is not true for the border survival probability.

Why, then, birds on the border tend to exchange position with neighbours less than what internal birds do?
First, and most trivially, we need to consider that border individuals - due to their peripheral location - lack neighbours on one side. Therefore, any movement larger than $l_\mathrm{B}$ but in the outward direction does not decrease the relative distance with any internal neighbour and leaves the bird on the boundary of the flock. To take this into account, the appropriate quantity to look at is therefore the probability that an individual moves less than $l_\mathrm{B}$ only in the inward direction. For a generic diffusion process there are not analytic expressions for this quantity. However, in the case of a Brownian walker the computation can be easily done \cite{gardiner}, leading to
\begin{equation}
P(t)={\mathrm{erfc}}(l_{\mathrm{B}}/\sqrt{2Dt}).
\label{brownian_surv}
\end{equation}
A fit of the data with this Brownian functional form is displayed in Fig.~4.  While this fit captures the convex shape of the curve, the empirical border survival probability systematically decays faster for times larger than the typical diffusive scale (\ref{taudiff}). This can be explained by noticing that even if border birds can arbitrarily move towards the outside still remaining peripheral, they in fact never increase too much their distance from the flock, as this would imply leaving the group and losing altogether the benefits of the collective motion. Better be in the border than gone astray~\cite{lima}. In getting Eq.~(\ref{brownian_surv}), however, all these large outer `walks' are considered as possible positive contributions to the permanence on the border at large times. Thus, we must expect the real survival probability (which does not include such walks) to be smaller than the one predicted by Eq.~(\ref{brownian_surv}), as we indeed find.

Finally, there might be an additional effect to be taken into account. Previous experimental observations show that the density of flocks is larger at the border than at the interior \cite{ballerini+al_08a}. This effect could be a consequence of the fact that birds compete with each other for a place in the interior of the flock. This struggle for the occupation of the same internal space would imply that when attempting to move inward, a border bird experiences an outward repulsion produced by its internal neighbours, pushing it outside again. The attempt to move inward is then reiterated, until by some fluctuation the bird successfully leaves the border. This mechanism clearly contributes to increase the survival probability of border birds as compared to internal ones.

\section{Discussion}

Our results show that diffusion in the centre of mass reference frame occurs with an exponent, $\alpha =1.73$, much larger than the Browninan case ($\alpha=1$). Birds within a flock are therefore strongly superdiffusive. How theoretical predictions of flocking diffusion compare with our data?
Hydrodynamic theories of flocking \cite{toner+tu_95,toner+tu_98,toner+tu+ulm_98} make some  predictions about the emergence of anomalous diffusion. In particular, in two dimensions these theories predict superdiffusive behaviour, with an exponent $\alpha=4/3$ \cite{toner+tu+ulm_98}. Numerical simulations in $2d$ models of self-propelled particles support these predictions \cite{toner+tu+ulm_98, gregoire+chate+tu_03}. However, these predictions have been made for two-dimensional  systems, whereas our data are in 3d. Hydrodynamic predictions in 3d are much harder to perform, but according to a conjecture put forward  in \cite{toner+tu_98} it would be expected $\alpha =1$ in $d=3$, in contrast with our result. On the other hand, numerical simulations in three dimensions \cite{chate+peruani_08}, give $\alpha=1.7$, in agreement with our experimental value. We believe that now that experimental data about diffusion are available, both theoretical and numerical studies in $3d$ should be reconsidered more carefully, as the prediction of the right diffusion properties can be a very effective model-selection tool.

Our diffusion data display strongly anisotropic behaviour. Motion is quite limited in the plane formed by flock velocity and gravity, while it is much stronger along a direction perpendicular to that plane. We can roughly identify this direction of maximal diffusion with the wings axis. There is a compelling geometric argument to explain the origin of anisotropic diffusion \cite{toner+tu_98}: if birds make small errors $\delta\theta$ in their direction of motion, their random displacement perpendicular to the mean direction of motion $\vec V$ is much larger than that along $\vec V$; the former is proportional to $\sin(\delta\theta)\sim \delta\theta$, while the latter is proportional to $1-\cos(\delta\theta)\sim \delta\theta^2 \ll \delta\theta$. Therefore, diffusion is suppressed along the direction of motion $\vec V$. This simple argument does not take into account the role of gravity, which has the effect of further depress vertical diffusion on the plane perpendicular to $\vec V$. As a consequence, one expects to have minimal diffusion along both $\vec V$ and gravity, and maximal diffusion along the direction perpendicular to them. This is exactly what we find.

When we consider mutual diffusion, namely how much a bird moves with respect to its nearest neighbours, we find diffusion
exponents similar to diffusion in the centre of mass reference frame, but much lower diffusion constant. In other words, birds move {\it less} with respect to their neighbours than with respect to the centre of mass. This fact is the consequence of the very strong and long-ranged spatial correlations of the velocity fluctuations observed in \cite{cavagna+al_10}. Neighbouring birds' displacements in the centre of mass reference frame are similar, so that birds do
not depart from each other as much as they move throughout the flock.

From the full individual trajectories we calculated the neighbours overlap $Q_M(t)$ and thus quantified how much, on average, the local neighbourhood
of a focal bird changes in time. Our data show that neighbours reshuffling occurs, so that each bird gradually changes all its interacting neighbours over time. There is no indication of a fixed structure of neighbours in the flock. In fact, we showed that a very simple model, whose only ingredient is mutual diffusion, reproduces quantitatively well the neighbours overlap, without the need of any extra dynamical ingredients. This fact seems to indicate that the neighbours each bird is interacting with at each instant of time are not selected on the basis of a biological criterium, but they just randomly happen to be there,
according to diffusion laws.

Even though neighbours reshuffling definitely occurs, it seems however {\it not} to be a very fast process.
To give full validity to such statement we should define a timescale (the birds' `clock'), which is not
straighforward. Still, we do expect any kind of update of the internal state of motion of a bird to happen on a rather fast time scale,
let us say definitely smaller than 0.1 seconds. Hence, the fact that, for example, it takes about 3.5 seconds to change only half of 10
neighbours (Fig.~3), really seems to indicate that neighbours permanence is rather high. This is interesting.
Indeed, according to several theoretical and numerical studies, the fact that the interaction network changes in time has the effect of reinforcing the alignment order in the flock \cite{toner+tu_95, chate_comment_07,ginelli_08}. Changing the neighbours over time amounts to have an {\it effective} number of interacting neighbours that is larger than the instantaneous one.

However, there may be a trade-off: exchanging neighbours {\it too} quickly could be detrimental for establishing long-range order in the flock.
At each time step one individual tries to align its velocity to that of its neighbours; but there is noise,
so that alignment is not perfect and it may take several time steps to consolidate consensus. If, however, the pool of neighbours changes
completely from one time step to the next, it will be very hard to beat noise and therefore to dynamically reach global consensus.
If a trade-off exists, there should be an optimal neighbours reshuffling rate that makes global-order easiest to achieve at the dynamical level.
However, even if an optimum exists, it does not imply that the natural system is actually at the optimum. The comparison of
theoretical models, where the rate of neighbours reshuffling can be artificially altered, with our experimental data, which give quantitative substance to
these speculations, can help understanding whether or not an optimum neighbours reshuffling exists and to what extent natural flocks of birds are close to such optimum.

Finally, we have investigated the dynamics of individuals at the border of the flock. What we find is an intriguing difference between motion within the flock and motion at the border. The survival probability of individuals at the border is indeed significantly larger than the survival probability of internal individuals:  birds stay on the border longer than the way internal birds keep their position inside the flock. Our analysis suggests that in doing this individuals on the border balance the tendency to exchange neighbours due to motion, the availability of void space outside the flock, and the resilience of internal neighbours to give up a more favourable position.

When a predator (like the Peregrin Falcon) attacks a flock, it is mostly birds on the border that gets captured. Hence, the
border is a dangerous place.  And yet, birds dynamics does not accelerate border turnover.
It seems that the flock self-organizes out of the individual selfish tendency not to stay at the border. This situation is reminiscent of the `selfish herd' scenario described by Hamilton \cite{hamilton_71}.
Border dynamics is very fascinating and very important, and we just started scratching the surface of it. New data, and more specifically  longer and more exhaustive trajectories, are needed to be able to fully unveil border dynamics.

\subsection*{Acknowledgements}
We thank Francesco Ginelli and John Toner for several interesting discussions. This work was supported in part by grants IIT--Seed Artswarm, ERC--StG n.257126, AFOSR--Z80910 and FP6-NEST 12682 STARFLAG. S.M.D.Q. is supported by a Marie Curie Intra-European Fellowship (FP7/2009) n. 250589.

\section{Methods}
\label{methods}

Analyzed data were obtained from experiments on large flocks of starlings ({\em Sturnus vulgaris}), in the field. Using stereometric
photography and computer vision techniques \cite{cavagna+al_08a,cavagna+al_08b} the individual 3D coordinates were measured
in groups of up to few thousands individuals \cite{ballerini+al_08b,cavagna+al_08a,ballerini+al_08a}. For a number of flocking
events (see Table 1), we could retrieve individual trajectories.  Each event consists of up to 40  consecutive 3D configurations
(individual positions), at time intervals of $0.1$s. We developed a tracking algorithm that connects the 3D spatial positions of the
{\it same} individual through time. Temporal matching between consecutive times is based on a patter-algorithm of the same kind as the
one used to solve the stereometric matching (see \cite{cavagna+al_08a}). This two-time match is effective but never complete.
At each instant of time a small percentage of individuals (typically below $5\%$ in our case) is not reconstructed due to occlusions on
the images and segmentation errors. Due to this, a mere iteration of two-time matches only brings a set
of very short interrupted trajectories. To overcome this problem, we developed a Monte Carlo algorithm that  allows for `ghosts' to simulate
the occurrence of missing 3D reconstructions, and  patches together pieces of trajectories by optimizing an appropriate measure combining
average smoothness, 3D constraints and number of ghosts. Thanks to this algorithm, we could retrieve a reasonable percentage of individual
trajectories as long as the entire event.

Given a flocking event, we considered the subset of retrieved long-lasting trajectories and computed the mean squared displacement and mutual square displacement, following
Eqs.~\ref{msd} and \ref{mutdiff}. To estimate the diffusion exponents and coefficients, we fitted the resulting time dependence in log-log scale between time lags of $0.4$ and $1.5$ seconds, which takes into account the length of all the data at our disposal. Results for the individual flocks are reported in Table 1, and correspond to super-diffusive behaviour. The statistical significance of this finding in connection with the finiteness of the time series was tested using synthetic data (see ESM for a full account of the procedure).

\appendix

\section*{Supplementary information} 

\subsection*{Mutual diffusion}

As stated in the main text, we computed the mutual diffusion as,
\begin{equation}
\delta r_{m}^2\left( t\right) \equiv \frac{1}{T-t}\frac{1}{N}%
\sum_{t_{0}=0}^{T-t-1}\sum\limits_{i=1}^{N}
\left[ |\vec{s}_{ij}\left(t_{0}+t\right)| - |\vec{s}_{ij}\left( t_{0}\right)| \right] ^{2}.
\end{equation}
where $\vec{s}_{ij}\left( t\right) \equiv $ $\vec{r}%
_{i}\left( t\right) -$ $\vec{r}_{j}\left( t\right) $ is the
position of bird $j$ (the nearest neighbour of $i$ at time $t_{0}$) in the reference frame of $i$.

\begin{figure}[tbh]
\begin{center}
\includegraphics[width=0.75
\columnwidth,angle=0]{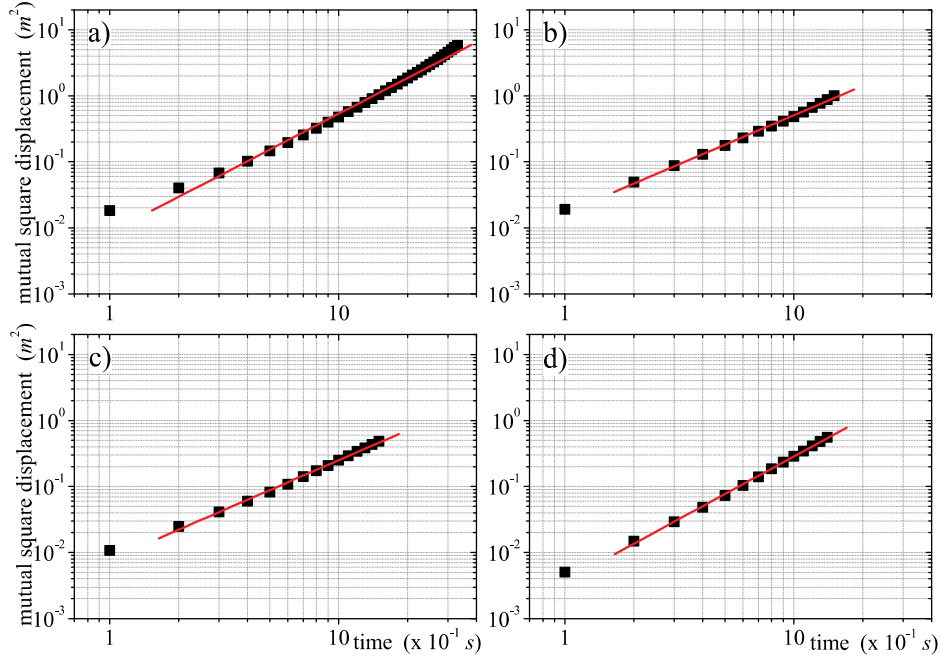}
\end{center}
\caption{Mutual diffusion of the nearest neighbour as defined in Eq.~(4) of the main text for 4 different flocking
events: a) 69-10; b) 48-17; c) 49-05; d) 28-10.}
\label{fig-2}
\end{figure}

Applying this formula to our trajectories, we obtained a typical behaviour depicted in Fig.~\ref{fig-2}, with averages values given in (6) of the main text. The values of $\alpha_m$ and $D_m$ for specific flocks are given in Table 1.

As commented in the main text, mutual diffusion is clearly suppressed with respect to diffusion in the centre of mass reference frame, due to the presence of correlations between birds. On the time scales accessible to our analysis, this is particularly evident in the values of the diffusion coefficients. Concerning the exponents, one has to be careful to draw conclusions. A linear regression between $\alpha$ and $\alpha_m$
 shows that the two exponents are correlated (correlation coefficient 0.95). Then, we used a paired Wilcoxon test to assess whether the two exponents are significantly different. More precisely, given the set of $\alpha$ values obtained for each individual flock and the set of $\alpha_m$ values obtained for each flock, the test quantifies whether the hypothesis that the medians of the two sets are equal can or cannot be rejected.
In our case, we find that the hypothesis cannot be rejected for a significance level of 0.05, but is rejected for a significance level of 0.1. In other terms, if we use a stronger significance criterium then - given our data - we cannot conclude that the exponents are different. If we opt for a slightly less stringent criterium then we would consider them as different. However, we stress that it was not our intention to make strong claims about the diversity of the exponents. Longer time lags and a better statistics would be necessary to this task. Our main point is that mutual diffusion is anyway smaller than diffusion in the centre of mass reference frame for the time lags at our disposal.

Concerning this point, since the exponent of the mean square displacement is related to asymptotic behaviour, one might wonder whether the similarity between the exponents $\alpha$ and $\alpha_m$ increases when considering only the last part of the diffusion curves.  For example, for one of the longest event that we have (69-10, panel a) in Fig~\ref{fig-2} ) we can notice an upper bending in the last part of the curve and it is reasonable to ask whether taking this bending into account would change some of the conclusions. Therefore,  we tried fitting the diffusion curves on the last interval $[1- 3.4]$ (to be compared with the interval $[0.4-1.5]$ used in Fig.~\ref{fig-2}). We get in this case $\alpha=1.66$ for diffusion and $\alpha_m=2.07$ for mutual diffusion. While these values are both larger than the ones obtained using the interval $[0.4-1.5]$ (see Table 1), they are nonetheless less similar (rather than more similar) to each other. We note in this respect that considering only the last part of the diffusion curves for the fit is rather risky, since for those points the statistics is smaller: an example of how this might fictitiously affect the retrieved exponent is given in Fig.~\ref{fig-si}. For most of other flocks the interval used in the paper $[0.4-1.5]$ is already rather short to be further reduced.

\subsection*{Neighbours overlap from diffusion}

The number $M$ of neighbors within a radius $R$ around some focal bird at some initial time $t_{0}$ is,
\begin{equation}
M \sim \rho R^{d}
\label{mt0}  \  ,
\end{equation}
where $\rho $ is the density of the system and $d=3$ is the dimension of space.
As time evolves from $t_{0}$ to $t_{0}+t$, the $M$ nearest neighbors move due to diffusion, occupying an expanded sphere of radius $R(t) > R$
around the focal bird. This means that the effective density $\rho(t)$ of these $M$ initial birds decreases,
\begin{equation}
\rho(t) =\frac{M}{R(t)^d} < \rho  \  .
\label{mt}
\end{equation}

\begin{figure}[tbh]
\begin{center}
\includegraphics[width=0.4
\columnwidth,angle=0]{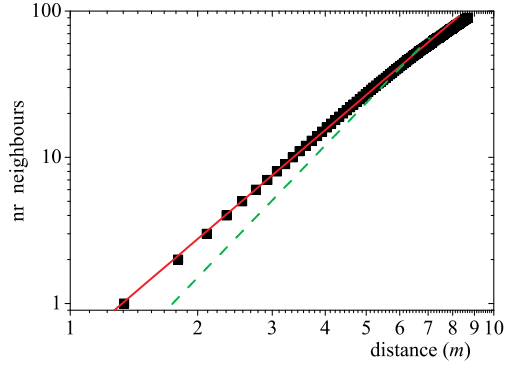}
\end{center}
\caption{Number of neighbors $M$ as a function of the radius $R$ of the sphere containing them.
We fit this data to the formula, $M = a R^{\hat d}$.
The full red line represents the best fit with fixed parameter $\hat d=3$. The dashed green line represents the best fit
where the exponent is let free. The result is $\hat d =2.3\pm 0.08$ [$R=0.9998$, $\protect\chi ^{2}=0.0057$, $p<0.0001
$]. Both fits are performed up to $M=50$.}
\label{zonko}
\end{figure}

At time $t_0+t$, the number $M(t)$, out of the $M$ initial birds, that {\it still} remain within a radius
$R$ (which is the distance that defines the $M$ nearest neighbors), is given by,
\begin{equation}
M(t)  =\rho(t) R^d\ = \frac{M}{R(t)^d} R^d \ .
\end{equation}
We can therefore work out the neighbours overlap,
\begin{equation}
 Q_{M}\left( t\right)   \equiv  \frac{M\left( t\right) }{M}=\frac{%
R^{d}}{R(t) ^{d}} \  .
\label{m3}
\end{equation}
We now make a very crude deterministic approximation and assume that the value of the radius $R(t)$ depends on the relative
diffusion of the birds, in the following way,
\begin{equation}
R(t)  \sim R +\sqrt{D_{ij}} \,t^{\alpha_{m}/2},
\label{r1}
\end{equation}
where $\alpha_{m}$ is the mutual diffusion exponent and $D_{m}$ is the mutual diffusion constant.
Substituting Eq.~(\ref{r1}) into Eq.~(\ref{m3}), we get,
\begin{equation}
Q_M(t)=\left(1+\sqrt{D_{m}} \; \frac{t^{\alpha_{m}/2}}{R}\right)^{-d}  \ .
\end{equation}
In real finite flocks, the dimension $d=3$ must be reduced to an effective value
$\hat d <3$ because of border effects, so that the formula, $M=4/3 \pi R^3 \; \rho$
gets modified to a more general expression, $M=a \;R^{\hat d}$, with $\hat d = 2.3$
and $a=0.5$ (see Fig.\ref{zonko}). From this we finally obtain,
\begin{equation}
Q_M(t)=\left(1+ c \; \frac{t^{\alpha_{m}/2}}{M^{1/\hat d}}\right)^{-\hat d} \ ,
\end{equation}
where $c=\sqrt{D_{m}} \, a^{1/\hat d}$. For flock 69-10, this expression gives $c=0.07$, while a fit of the
data in Fig.4 of the main text  gives $c=0.05$. Considering the crude approximation
that we are using, eq.\eqref{r1}, these two numbers are reasonably close to each other.

\subsection*{Determination of the principal axis of diffusion}

To determine the main axis of diffusion we consider a matrix generalization of  Eq.~(3),
\begin{equation}
\Delta_{\mu \nu }\left( t\right) \equiv \frac{1}{%
T-t}\frac{1}{N}\sum_{t_{0}=0}^{T-t-1}\sum\limits_{i=1}^{N}\left[ r_{i,\mu
}\left( t_{0}+t\right) -r_{i,\mu }\left( t_{0}\right) \right] \left[ r_{i,\nu
}\left( t_{0}+t\right) -r_{i,\nu }\left( t_{0}\right) \right],
\label{covmatrix}
\end{equation}
where $r_{i,\mu }\left( t\right) $ represents the $\mu \in \left\{x,y,z\right\} $ Cartesian component of the position of bird $i$ with respect to the center of mass at time $t$ (and, in the same way, $\nu \in \left\{x,y,z\right\}$ indicates another component). The standard mean square displacement $\delta r^2$ defined in the main text is simply the trace of this matrix, i.e. $\delta r^2 = \sum_\mu \Delta _{\mu \mu }$.

We now want to compute the eigenvalues and eigenvectors of this diffusion matrix, for any given time lag $t$. The eigenvalues $\left\{ \lambda _{1}(t),\lambda _{2}(t),\lambda _{3}(t)\right\} $ of $\mathbf{\Delta}$ are given - at any time $t$ - by the solutions of the equation%
\begin{equation}
\det \left\vert \mathbf{\Delta}(t)-\lambda(t) \,\mathbf{I}\right\vert =0,
\end{equation}
where $\mathbf{I}$ is the identity matrix and $\det \left\vert \mathbf{C} \right\vert $ represents the
determinant of the matrix $\mathbf{C}$.
Each of the $\lambda _{a}$ values (with $a=1\cdots 3$)  can be associated with a vector $%
\mathbf{w}_{a }$ such that,%
\begin{equation}
\mathbf{\Delta\,}\mathbf{w}_{a }-\lambda _{a} \,\mathbf{I\,}\mathbf{w}_{a}=0,
\end{equation}
with every $\mathbf{w}_{a }$ orthogonal to the others.  In consequence, if $\lambda _{1}>\lambda _{2}>\lambda _{3}$, we are able to establish three independent (orthogonal) directions defined by the unitary eigenvectors, $\mathbf{u} _{a } \equiv \frac{\mathbf{w}_{a}}{\left\Vert \mathbf{w}_{a }\right\Vert }$, where $\mathbf{u}_{1}$ describes the direction of maximum diffusion, $ \mathbf{u}_{2}$ the direction of second maximum diffusion and  $\mathbf{u}_{3}$ the direction of minimum diffusion.

\begin{figure}[tbh]
\begin{center}
\includegraphics[width=0.75
\columnwidth,angle=0]{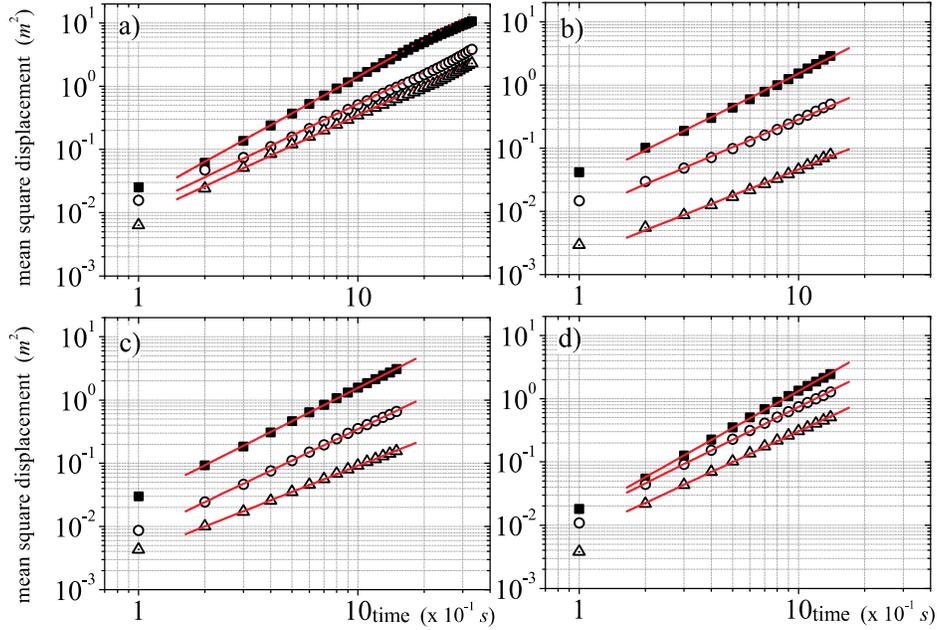}
\end{center}
\caption{Mean square displacement along the $3$ principal axes of diffusion. Same flocks labels as in the previous figures.}
\label{aniso}
\end{figure}

Since the trace (sum of the diagonal entries) of any matrix is invariant we therefore have that the standard mean square displacement  (the trace of the diffusion matrix) is given by the sum of  eigenvalues, i.e. $\delta r^2(t) = \sum_\mu \Delta _{\mu \mu }(t)= \lambda _{1}(t)+\lambda_{2}(t)+\lambda _{3}(t)$.
The standard mean-square displacement can therefore be decomposed as the sum of the displacements arising along the three principal axis, each one given by the corresponding eigenvalue as a function of time. In Fig.~\ref{aniso} we report the behaviour of the mean-square displacement along the three principal axis as a function of time, for $4$ flocking events. One can clearly see that the corresponding exponents (slopes of the curves) are different along the three axis, meaning that diffusion is anisotropic. The values of the diffusion exponents and diffusion coefficients along the three axis, averaged over all flocking events, are reported in the main text and confirm this conclusion.

\subsection*{Statistical significance}
\label{significance}

As briefly discussed in the main text, given a flocking event, we considered the subset of retrieved long-lasting trajectories and for each  frame we calculated the center of mass coordinates $R_{CM}$. Then, we computed the mean squared displacement and mutual square displacement, following Eqs (3) and (5) of the main text.  We note that - for what concerns diffusion  - we could have used a larger sample of trajectories: when computing the mean-square displacement on a time lag equal to $t$ we can indeed consider all trajectories that are long at least as $t$ (and not only the long-lasting ones). Results do not change much and one would get very similar exponents.

To estimate the diffusion exponents, we fitted the resulting time dependence in log-log scale. The results in Table 1 correspond to averages of the numerical adjustments between time lags of $0.4$ and $1.5$ seconds, which takes into account the length of all the data at
our disposal. The exponents that we find are much larger than the value $\alpha=1$, indicating that flocks do exhibit super rather than standard diffusion. To check that this finding is not an artifact due to the finiteness of the time series (which causes a small number of samples as the lag approaches the series length), and assess the statistical significance of our results, we produced synthetic data obeying standard diffusion on the same time lags as our data. In this way, we verified that the exponents that we find for real flocks are consistently greater than the exponents corresponding to the percentile 95 obtained for normal diffusion series. Indeed, for the series
we have studied, the critical values associated with this percentile have its maximal value equal to $1.64$.

Let us now explain in detail the procedure that we followed.
In order to analyze the statistical hypothesis of Brownian motion as well as
the determination of the critical diffusion exponent, $\alpha ^{\ast }$, the
percentile $95$ of which corresponds to the value obtained by numerical
adjustment of a time series of length $T$, we have carried out the
generation of long series from which a patch of length $T$ taken. The
diffusion, $\delta r^2\left( t\right) $, is analytically obtained from,%
\[
\delta r^2 \left( t\right) =\int \int \mathcal{C}_{\mathbf{v}}\left( \tau
^{\prime },\tau \right) \,d\tau \,d\tau ^{\prime },
\]%
where the covariance of the velocities $\mathcal{C}_{\mathbf{v}}$ is defined as in Eq.~(12) in the main text, and with $\tau ^{\prime }=\tau +t$.
If correlations decay in time as power-law $%
\mathcal{C}_{\mathbf{v}}\left( \tau +t,\tau \right) \sim t^{-\xi }$, then
the diffusion is a power function with respect to $t$, $\delta r^2 \left(
t\right) \sim t^{2-\xi }$, and thus $\alpha =2-\xi $. To generate power-law
correlated velocities in a savvy way we have resorted to the Wiener-Khinchin
theorem relating the correlation function and the spectral density and
proceeded as follows;

\begin{itemize}
\item Generate a series of Gaussian time series, $\left\{ u\right\} $, of
length $N$ (with $N$ being a odd number, \emph{e.g.}, $10^{6}+1$);\footnote{%
The odd number is a simple trick to avoid the frequency $0$ which is
associated with a singularity of the power spectrum.}

\item Compute its Fourier transform, where the element $u_{i}$ corresponds
to a value $\tilde{u}\left( f=\frac{i-1}{N}-\frac{1}{2}\right) $;

\item Set apart the absolute value and multiply $\exp \left[ \mathrm{i}%
\,\arg (\tilde{u}\left( f\right)) \right] $ by the square root of the Fourier
transform of $\mathcal{C}_{\mathbf{v}}\left( \tau +t,\tau \right) $, which
in this case is a power-law function as well, $S\left( f\right) \sim
\left\vert f\right\vert ^{-\xi -1}$;

\begin{figure}[tbh]
\begin{center}
\includegraphics[width=0.45\columnwidth,angle=0]{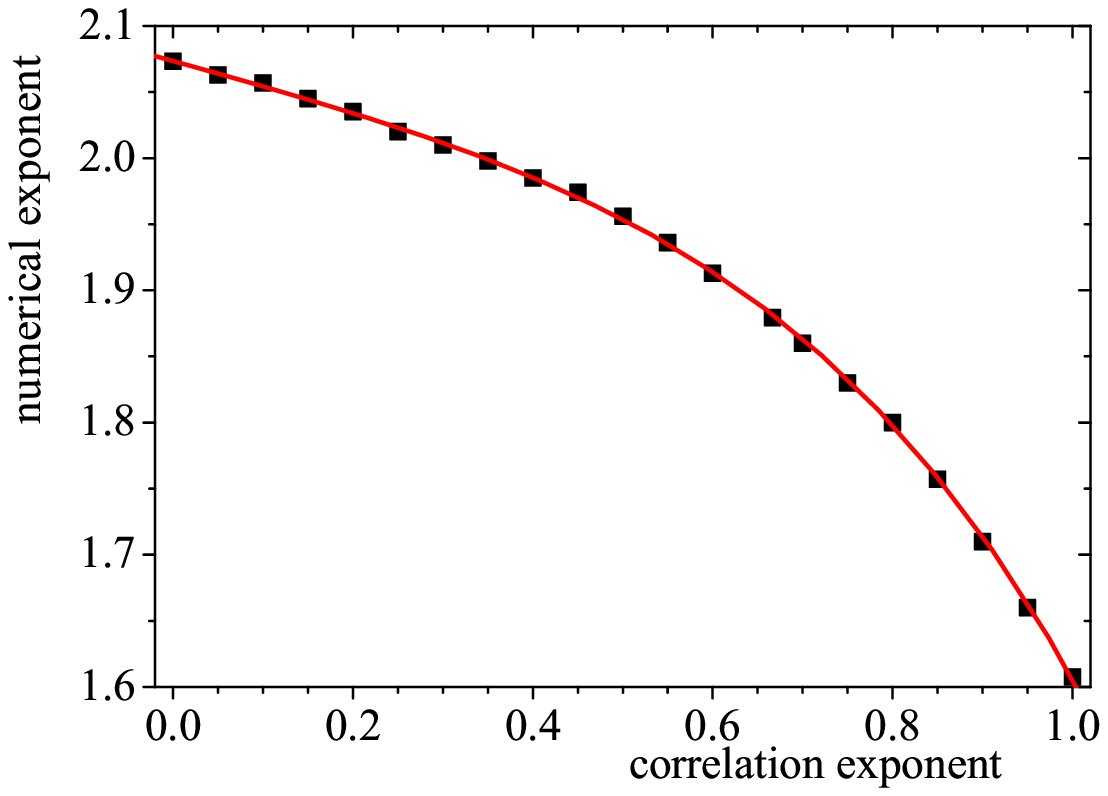}
\includegraphics[width=0.45\columnwidth,angle=0]{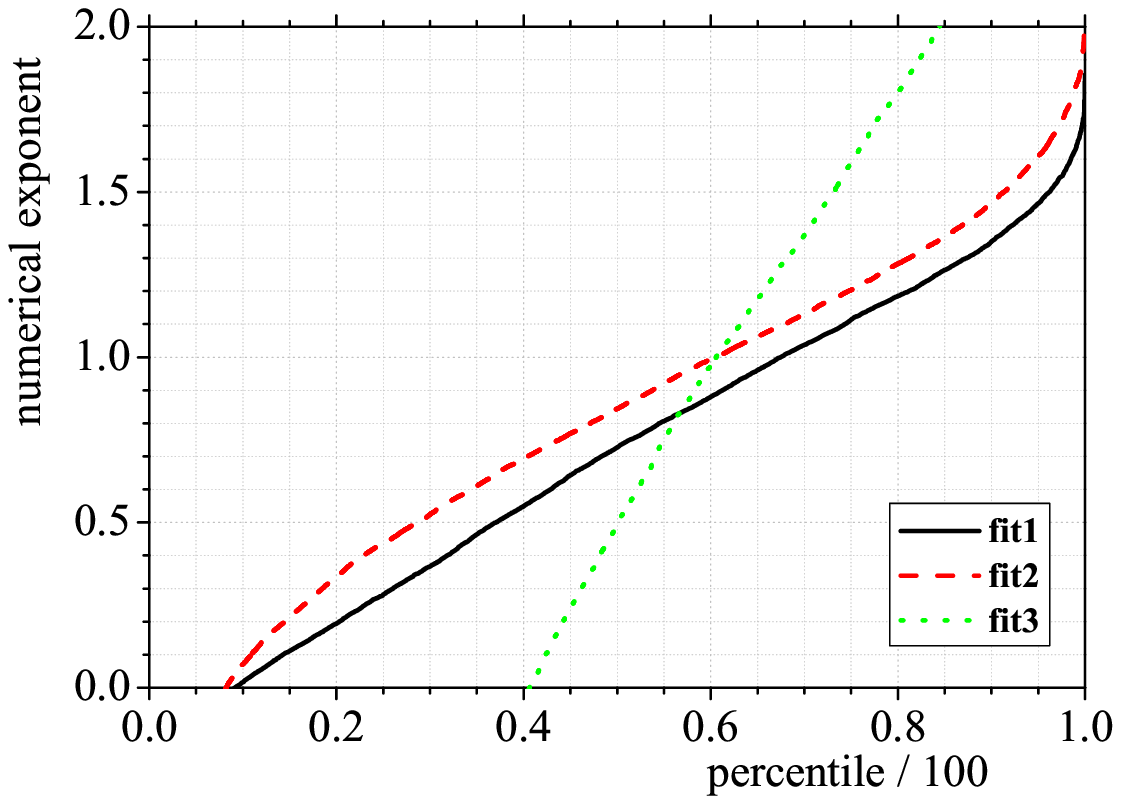}
\end{center}
\caption{
Left panel:  exponent obtained by the numerical adjustment of the diffusion,
which has been computed in patches of Brownian motion with the same length
of the flock 69--10 and averaged over the $500$ samples (one sample
cycle), as a function of the percentile (divided by $100$). The line \textbf{fit1}
takes into account lags from $1$ until $33$, \textbf{fit2} corresponds to the
interval used in the manuscript and \textbf{fit3} is obtained considering the
interval of lags between $16$ and $33$. Right panel: The percentile $95$ as
a function of the correlation exponent for the dataset 69--10 and
considering the lag interval \textbf{fit2}. The points have been obtained from $10^{4}$
sampling cycles and the line has been obtained by interpolation using the
points.}
\label{fig-si}
\end{figure}

\item Invert the Fourier transform and multiply the outcome by $\left(
-1\right) ^{i+1}$ to finally obtain the correlated series $v\left( i\right) $.
\end{itemize}

Afterwards, the series is summed to create position sequences. We have
considered values of $\xi $ between $0$ and $1$, which after integration
bear ballistic and Brownian trajectories, respectively.

As an illustration, in Fig.~\ref{fig-si}, we present the exponent obtained by the
numerical adjustment of the diffusion, which has been computed in patches of
Brownian motion with the same length of the flock 69--10 and averaged over
the same number of birds, as a function of the percentile and the exponent
corresponding to the percentile $95$ as a function of the correlation
exponent, $\xi $.

We have also considered the hypothesis that velocities follow a Langevin stochastic equation with a typical scale equal to $k ^{-1}$,
the correlation function of which decays in the form of an exponential. This case also leads to a functional dependence diffusion given by
$k \, t + \exp [- k\, t] -1$, which that does fit for our field values.

\subsection*{Border definition and Border diffusion}

Given a flock, in order to compute several global and statistical quantities it is necessary to appropriately define its exterior border (see \cite{cavagna+al_08b} for a thorough discussion of this problem).
Flocks are typically non-convex systems. Therefore, standard methods to define the border, like the convex-hull, are inadequate because they are unable to detect concavities. To overcome this problem, we used the so-called  `$\alpha$-shape algorithm \cite{edelsbrunner_94}. The main idea of this method is the following: given  a set of 3D points,  one `excavates' the set of points with spheres of radius $\alpha$, so that all concavities of size larger than $\alpha$ are detected. Formally, one selects the sub-complex of the Delaunay triangulation on scale $\alpha$ (the $\alpha$-complex) and the external surface of this triangulation defines the border.

The scale $\alpha$ must be appropriately chosen. If $\alpha$ is too large, some concavities are neglected and void regions are included as being part of the flock. Too small values of $\alpha$, on the other hand, might cause the sphere to penetrate the flock and break it into sub-connected components. A robust criterion is to look at the density of the internal points as a function of $\alpha$ \cite{cavagna+al_08b}\cite{tavarone_09}. This quantity typically has a maximum, which defines a natural scale for $\alpha$.

For all the analyzed flocking events, the border has been computed following the above procedure. We note that, since flocks change shape in time, the border must be computed and re-defined at each instant of time. Besides, due to the continuous movement of individuals through the group, the individuals belonging to the border change from time to time.

Once we have defined the border, we can ask whether the diffusion properties that we have described and quantified in the main text are similar for birds belonging to the border and birds well inside the bulk of the group. Let us consider for example the behaviour of the mean-square displacement. Clearly, since individuals on the border do not remain there forever, but at some point leave the boundary, on large time scales we cannot even distinguish border and internal birds. However, on shorter time lags we can try to investigate such a difference.

\begin{figure}[tbh]
\begin{center}
\includegraphics[width=0.75\columnwidth,angle=0]{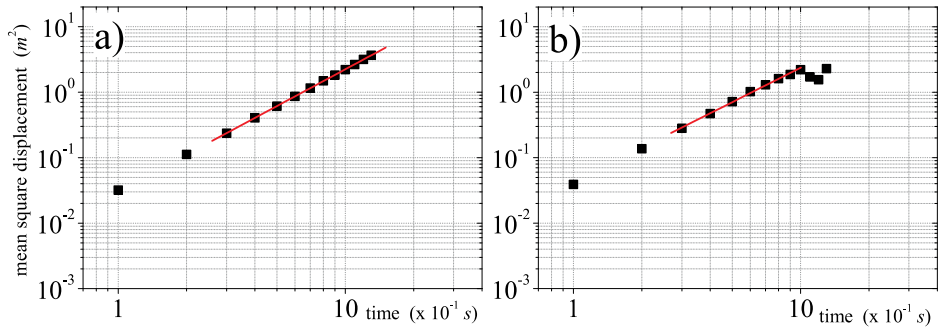}
\end{center}
\caption{Mean square displacement in the centre of mass reference frame for a) internal birds and b) birds on the border of the flock. In panel b) only individuals that remain on the border for the entire time lag $t$ are considered when computing $\delta r^2(t)$ (see text). The exponents are $\alpha=1.86 \pm 0.02$ for internal birds, and $\alpha=1.78 \pm 0.03$ for border birds, to be compared with the value $\alpha=1.83\pm0.01$ obtained with all the birds. For the diffusion coefficients we get (values divided for $10^{-2}$ as in Table 1) $D= 3.1 \pm 0.03$ (internal) and $D=4.2 \pm 0.05$ (border), to be compared with $D=3.8\pm0.1$ (all birds).}
\label{border_diffusion}
\end{figure}

\begin{figure}[tbh]
\begin{center}
\includegraphics[width=0.75\columnwidth,angle=0]{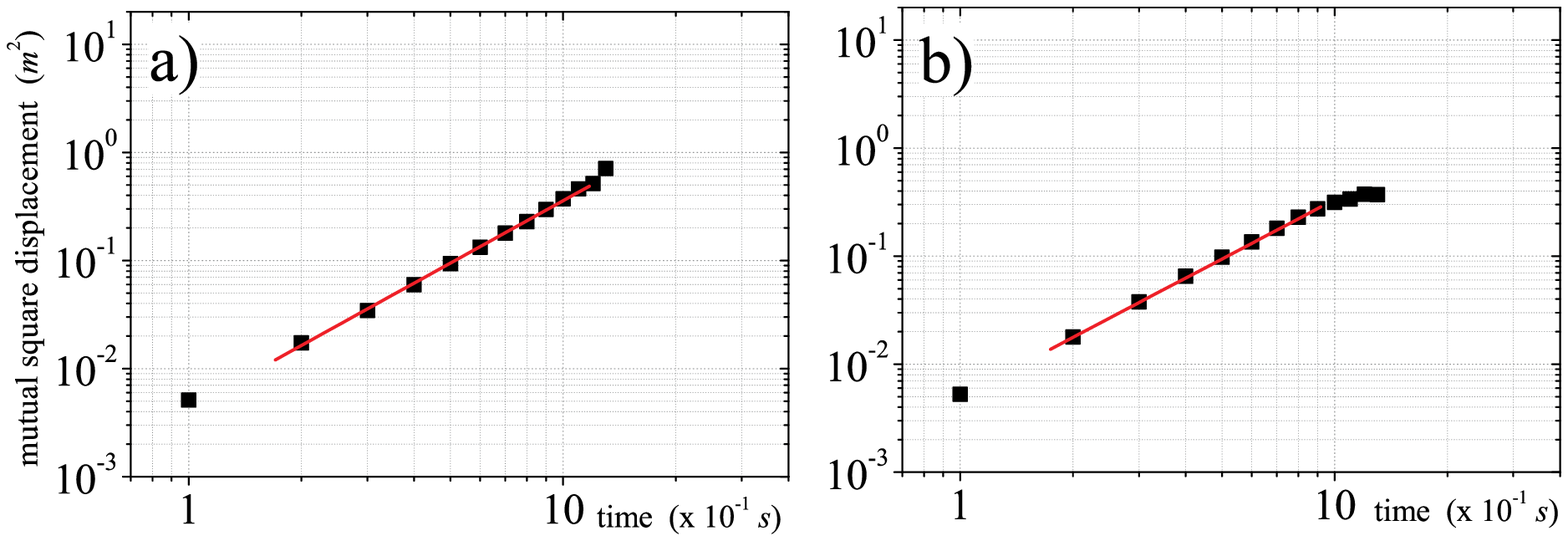}
\end{center}
\caption{Mean square displacement relative to mutual diffusion for a) internal birds and b) birds on the border of the flock. In panel b) only pairs of individuals (a bird and its nearest-neighbour) that remain on the border for the entire time lag $t$ are considered when computing $\delta r_m^2(t)$ (see text). The exponents are $\alpha_m=1.92\pm0.02$ for internal birds, and $\alpha_m=1.84\pm0.04$ for border birds, to be compared with the value $\alpha_m=1.88\pm 0.02$ obtained with all the birds. Diffusion coefficients are (values divided for $10^{-2}$ as in Table 1) $D= 0.42 \pm 0.06$ (internal) and $D=0.35 \pm 0.03$ (border), to be compared with $D=0.37\pm 0.04$ (all birds). }
\label{border_mutualdiffusion}
\end{figure}

To do this, for any time-lag $t$ we divided the statistical sample (segments of trajectories that are long $t$ steps) into trajectories where the bird belonged to the border for the whole period $t$, and internal birds (the complementary set). We then computed the mean-square displacement as a function of $t$. For statistical reasons we did this on the largest flock that we have (event 28-10). The results are shown in Fig.~\ref{border_diffusion} for diffusion in the centre of mass, and in Fig.~\ref{border_mutualdiffusion} for mutual diffusion.

As one can see from these figure, the exponents of internal diffusion are very similar to the ones computed with all the trajectories. Both the exponents and the diffusion coefficients of the boundary birds are slightly smaller (even if not too much). We note that - due to the procedure we used - the results for mutual diffusion on the border refer to individuals who {\it both} are on the border for the time lag $t$. Thus, they describe how a bird on the border moves with respect to its nearest-neighbour {\it on the border}. However, they do not provide any information on how birds on the border move with respect to their first {\it internal} nearest-neighbour. We tried also to investigate internal-border diffusion, but the statistics is really poor to draw any conclusions. For this reason we preferred to look at the survival probability of the permanence on the border (see main text), which is statistically more robust.

As already underlined, the above measurements of border diffusion suffer from a very much reduced statistical sample, especially at the larger times (only individuals that remained on the border up to $t$ are included). For smaller flocks this make the above analysis not feasible.

\subsection*{Computing the survival probability for internal individuals}

As discussed in the main text, we can define the survival probability of internal individuals  as the probability that in the centre of mass reference frame a bird has moved less than a distance $l_\mathrm{B}$ in a time $t$, where $l_\mathrm{B}$ is the typical distance of the nearest-neighbour. This probability gives an estimate of how much time is needed for a bird to exchange location with respect to the first shell of neighbours, and is a relevant benchmark to be compared with the border survival probability.

We know that birds at the interior of the flock obey super-diffusive behaviour. Thus, we would like to compute the survival probability starting from the diffusion properties of the birds. In the case of a generic diffusion exponent $\alpha$ there are not analytic exact computations. However, there are some rigorous bounds for the short and large time limits of the survival probability \cite{monrad+rootzen_95,metzler+klafter_00}. More precisely, the theory predicts that
\begin{equation}
P(t) \sim \exp \left[- \frac{k(t)}{(l_\mathrm{B}^2/D)^{1/\alpha}}  \right] \  ,
\label{small2}
\end{equation}
where $k(t)$ is an unknown function, for which however we know that $k(t)=k_1 t$ in the short time regime, and $k(t)=k_2 t$ in the long time regime.
For Brownian diffusion ($\alpha=1$), one has $k_1=k_2$ and this function becomes a simple exponential. For super-diffusion ($\alpha>1$) one has $k_2>k_1$, i.e. there is an initial exponential decay at short times and a faster decay in the large time regime. The survival probability for internal individuals that we computed on the data is qualitatively consistent with this $\alpha>1$ prediction. Besides, if we fit the initial part of the curve (first 8 points) with an exponential function of the form (\ref{small2}) with $k(t)=k_1 t$, and we use the experimental values for $\alpha$ and $D$ in each flock, we find that the constant $k_1$ is fairly consistent in all flocks, and equal to  $1.42 \pm 0.15$. In other terms, the short time regime of the survival probability for internal birds is well described by Eq.~(\ref{small2}) with a flock independent constant $k_1$.

\end{document}